%% file: ttbar_Afb_PRD_arXiv2.tex
\documentclass[aps,prd,twocolumn,showpacs,superscriptaddress,amsfonts,floatfix]{revtex4-1}
\usepackage{amsmath}
\usepackage{graphicx}
\usepackage{bm}
\usepackage{epsfig}
\usepackage{psfrag}
\usepackage{multirow}
\usepackage{dcolumn}
\usepackage{amssymb}

\newcommand {\plainafb}	{\ensuremath{A_{\mathrm{FB}}}}
\newcommand {\afb}	{{\plainafb}}

\newcommand {\pbar}	{{\ensuremath{\bar p}}}
\newcommand {\ppbar}	{{\ensuremath{p\pbar}}}
\newcommand {\tbar}     {{\ensuremath{\bar t}}}
\newcommand {\ttbar}    {{\ensuremath{t\tbar}}}
\newcommand {\qbar}     {{\ensuremath{\bar q}}}

\newcommand {\itW}      {\ensuremath{W}}
\newcommand {\pjets}    {{\textrm{+jets}}}
\newcommand {\wpj}      {{\ensuremath{W\!+}jets}}
\newcommand {\wpjmath}  {{W\!+\mathrm{jets}}}
\newcommand {\leptonpj} {{lepton+jets}}
\newcommand {\lpj}      {{\ensuremath{l\pjets}}}
\newcommand {\epj}      {{\ensuremath{e\pjets}}}
\newcommand {\mpj}      {{\ensuremath{\mu\pjets}}}

\newcommand {\GeV}        {{\ensuremath{\,\mathrm {GeV}}}}

\newcommand {\ifb}        {{\ensuremath{\,\mathrm {fb}^{-1}}}}
\newcommand {\Tesla}        {{\ensuremath{\,\mathrm {T}}}}

\newcommand{\lep}         {\ensuremath{l}} 
\newcommand {\pt}         {{\ensuremath{p_T}}}
\newcommand {\ylep}       {{\ensuremath{y_\lep}}}
\newcommand {\qlep}       {{\ensuremath{q_\lep}}}
\newcommand {\qyl}        {{\ensuremath{\qlep\ylep}}}
\newcommand {\absyl}      {{\ensuremath{\left|\ylep\right|}}}
\newcommand {\yth}        {{\ensuremath{y_{t,\textrm{had}}}}}
\newcommand {\ytl}        {{\ensuremath{y_{t,\textrm{lep}}}}} 
\newcommand {\dy}         {{\ensuremath{\Delta y}}}
\newcommand {\absdy}      {{\ensuremath{\left|\dy\right|}}}

\newcommand {\ttpt} {{\ensuremath{p_T^\ttbar}}}

\newcommand{\emiss} {{/\!\!\!\!E}}
\newcommand{\met}   {{\ensuremath{\emiss_T}}}
\newcommand{\mex}   {{\ensuremath{\emiss_x}}}
\newcommand{\mey}   {{\ensuremath{\emiss_y}}}
\newcommand{\lbpt}  {{\ensuremath{p^{\mathrm{LB}}_{T}}}}
\newcommand{\ktmin} {{\ensuremath{k^{\min}_T}}}
\newcommand{\mjj}   {{\ensuremath{M_{jj}}}}
\newcommand{\chisq}  {{\ensuremath{\chi^2}}}

\newcommand {\afbl}   {{\ensuremath{\plainafb^\lep}}}
\newcommand {\mttbar} {{\ensuremath{m_\ttbar}}}

\newcommand {\Ntt}    {{\ensuremath{N_{\ttbar}}}}
\newcommand {\Nw}     {{\ensuremath{N_{\wpjmath}}}}
\newcommand {\Nmj}    {{\ensuremath{N_{\mathrm{MJ}}}}}
\newcommand {\Nf}     {{\ensuremath{N_{\mathrm{F}}}}}
\newcommand {\Nb}     {{\ensuremath{N_{\mathrm{B}}}}}
\newcommand {\Nfl}    {{\ensuremath{N_{\mathrm{F}}^\lep}}}
\newcommand {\Nbl}    {{\ensuremath{N_{\mathrm{B}}^\lep}}}
\newcommand {\Nfwd}   {\Nf}
\newcommand {\Nbwd}   {\Nb}
\newcommand {\Nlfwd}  {\Nfl}
\newcommand {\Nlbwd}  {\Nbl}

\newcommand {\DZ}     {{D0}} 

 \newcommand {\pythia}   {{\sc pythia}}
 \newcommand {\alpgen}   {{\sc alpgen}}
 \newcommand {\mcatnlo}  {{\sc mc@nlo}}
 \newcommand {\herwig}   {{\sc herwig}}
 \newcommand {\tunfold}   {{\sc tunfold}}

\newcommand {\etal}     {\emph{et al.}}

\newcommand {\tablestrut} {\rule{0pt}{2.5ex}}

\DeclareMathOperator{\sgn}{sgn}
\begin{document}

\hspace{5.2in} \mbox{FERMILAB-PUB-11-347-E}

\title{Forward-backward asymmetry in top quark-antiquark production}

\input author_list_for_ttbar_afb.tex

\date{July 25, 2011}
%
%
%
\begin{abstract}
We present a measurement of forward-backward asymmetry
in top quark-antiquark production in proton-antiproton 
collisions in the final state containing a lepton and at least four jets.
Using a dataset corresponding to an integrated luminosity of $5.4\ifb$, 
collected by the \DZ\ experiment at the Fermilab Tevatron Collider,
we measure the \ttbar\ forward-backward asymmetry to be $\left(9.2 \pm 3.7 \right)$\%
at the reconstruction level.
When corrected for detector acceptance and resolution, the asymmetry is found 
to be $\left(19.6 \pm 6.5\right)$\%. 
We also measure a corrected asymmetry based on the lepton from a top quark decay,
found to be $\left(15.2 \pm 4.0\right)$\%. 
The results are compared to predictions based on the next-to-leading-order QCD generator \mcatnlo. 
The sensitivity of the measured and predicted asymmetries to the modeling of gluon radiation is discussed.

\end{abstract}


\maketitle
%
%
\section{Introduction}
The top quark is the heaviest observed elementary particle.
As the only fermion whose mass is close to the electroweak scale, 
it may play a special role in electroweak symmetry breaking.  
So far, the measured top quark production and decay properties are consistent with predictions of the standard model (SM).
Although the top quark was discovered more than 15 years ago~\cite{bib:Abachi:1995iq, bib:Abe:1995hr}, 
the precision of many of these measurements
is still limited by sample size~\cite{bib:review},
and more precise measurements may yet uncover evidence for processes beyond the SM that contain top quarks.

Quantum chromodynamics (QCD) predicts that top quark-antiquark (\ttbar) production in quark and antiquark collisions is
forward-backward symmetric at leading order (LO). 
However, a positive asymmetry appears at higher orders. 
The asymmetry is such that the top quark is preferentially emitted in the 
direction of the incoming light quark, 
while the antitop quark follows the direction of the incoming antiquark~\cite{bib:asymwhy}. 
At the Tevatron, interactions between valence quarks 
dominate \ttbar\ production, so that the 
direction of the incoming quark almost always coincides with that of the proton. 
Thus, the Tevatron is well suited to studying such asymmetry. 

Processes beyond the SM can modify the \ttbar\ production asymmetry 
if, for example, axial currents contribute to $s$-channel 
production~\cite{bib:axigluon}, or if there is an abnormal enhancement of $t$-channel 
production~\cite{bib:tchannel}. 
In \DZ's previous study of this asymmetry~\cite{bib:p17PRL},
we set limits on the fraction of \ttbar\ events
produced via a new, heavy, mediating particle in the $s$ channel. 

After analyzing datasets corresponding to about 1\ifb\ of integrated luminosity each, 
the \DZ\ and CDF Collaborations found positive asymmetries that were consistent
with next-to-leading order (NLO) predictions~\cite{bib:p17PRL, bib:CDFPRL}. 
The CDF Collaboration recently reported several results based on a dataset 
corresponding to an integrated luminosity of $5.3\ifb$~\cite{bib:CDFdep}. 
The asymmetry in one particular subset of CDF data differs
by more than three standard deviations (SD) from the NLO prediction.

In this article we report a new study of forward-backward 
asymmetry in \ttbar\ production using a dataset corresponding to an integrated luminosity of $5.4\ifb$,
collected by the \DZ\ experiment. 
We define the asymmetry in terms of
the rapidity difference between the top and antitop quarks.
The rapidity $y$ is defined as $y\left(\theta,\beta\right)=\frac{1}{2}\ln\left[\left(1+\beta\cos\theta\right)\right.$ 
$\left./\left(1-\beta\cos\theta\right)\right]$,
where $\theta$ is the polar angle
and $\beta$ is ratio of a particle's momentum to its energy.
\DZ\ uses a cylindrical coordinate system, with the $z$-axis pointing
along the direction of the proton beam, and $\phi$ defined as the azimuthal angle.
We employ a kinematic fitting technique
to fully reconstruct the \ttbar\ candidate events. 
The results of the kinematic fit are used to measure the reconstructed \ttbar\ asymmetry.
We then correct for acceptance and detector resolution to find the inclusive production asymmetry.

We also present an asymmetry based on the rapidity and charge of 
the electron or muon from a top quark decay~\cite{bib:asymtev}. 
This method is less dependent on detector resolution than full \ttbar\ event
reconstruction and is sensitive to the underlying production asymmetry, 
thus providing a valuable cross check. The lepton-based asymmetry
is also directly sensitive to the polarization of the top quarks, 
and may be larger than the top quark asymmetry in some
new physics scenarios~\cite{bib:polarizedview}.

Finally, we discuss the predicted dependence of the asymmetry on gluon radiation. 
We verify the modeling of this radiation using the transverse momentum of the \ttbar\ system.

%
%
\section{\DZ\ Detector}
\label{sec:detector}
\DZ\ is a multipurpose detector designed to identify leptons, photons, and jets. 
 The central tracking system, consisting of a 
silicon microstrip tracker and a central fiber tracker, is  
located within a 1.9\Tesla\ superconducting solenoidal 
magnet~\cite{bib:run2det}. Tracks of charged particles can be reconstructed
for pseudorapidities $|\eta|<2.5$.
Central and forward preshower detectors are positioned in front of the calorimeter cryostats. 
Electrons, photons, and hadronic jets are identified using
a liquid-argon and uranium calorimeter, which has a 
central section covering $|\eta|$ up to 
$\approx 1.1$, and two end sections that extend coverage 
to $|\eta|\approx 4.2$~\cite{bib:run1det}. 
Muons are identified within $|\eta|<2$, 
using a muon system consisting of a layer of tracking detectors and scintillation
counters located in front of 1.8\Tesla\ iron toroids, followed by two similar layers 
after the toroids~\cite{bib:run2muon}. The luminosity is measured using plastic scintillator 
arrays placed in front of the endcap calorimeter cryostats.

To identify $b$ jets, we construct a neural network that combines variables characterizing 
the properties of secondary vertices and of tracks with large impact parameters
relative to the primary \ppbar\ interaction vertex (PV)~\cite{bib:btagging}.
The $b$-tagging requirement used in this analysis has an efficiency of about 70\%
for identifying $b$ jets originating from top quark decay,
and a misidentification probability of about 8\% for light flavored jets associated with the production of $W$ bosons.

%
%
\section{Event Selection and Reconstruction}
\label{sec:selection}
We select $\ttbar(X)\to W^{+}bW^{-}\bar{b} (X)$ events,  where one $W$ boson decays 
to $q\qbar'$ and the other decays to $l\bar{\nu}_{l}$. 
We select electrons and muons, which may arise directly from the $W\to l\bar{\nu}_{l}$ 
decay or through an intermediate $\tau$ lepton.
This \ttbar\ decay chain is referred to as the \lpj\ channel. 

The experimental signature of the \lpj\ channel is one isolated lepton with 
large transverse momentum (\pt), a significant imbalance in transverse momentum (\met) from 
the undetected neutrino, and four or more jets: two from 
the $W\to q \qbar'$ decay and the other two from fragmentation of the $b$ quarks.
We refer to the top quark that decayed to $b q \qbar'$ as the ``hadronic'' 
top and to the other top quark as the ``leptonic'' top.
Either of these terms can refer to the top quark or the antitop quark.
The electric charge of the lepton identifies the electric charge of the leptonic top.
The hadronic top is assumed to have the opposite charge. 

%
%
\subsection{Event Selection}
The event selection criteria used in this article are similar to those used to measure 
the \ttbar\ production cross section in the \lpj\ channel~\cite{bib:D0xsect}.
The reconstruction and identification of jets, leptons, and \met\ is described
in Ref.~\cite{bib:objid}.
The \epj\ and \mpj\ channels have similar event selection requirements.  
Events are triggered by requiring either a lepton ($e$ or $\mu$) or a lepton and a jet.
To select \epj\ events we require:
\begin{itemize}
\item one isolated electron with $\pt>20\GeV$ and $|\eta| < 1.1$,
\item $\met>20\GeV$, and
\item $\Delta\phi(e,\met) > \left(2.2 - 0.045 \cdot \met / \mathrm{GeV} \right)$ radians.
\end{itemize}
For \mpj\ events, we impose the following criteria:
\begin{itemize}
\item one isolated muon with $\pt>20\GeV$ and $|\eta| < 2.0$,
\item $25\GeV < \met<250\GeV$,
\item $\Delta\phi(\mu,\met) > \left(2.1 - 0.035 \cdot \met  / \mathrm{GeV} \right)$ radians, and
\item $(p_T^\mu+\met)^2-(p_x^\mu+\mex)^2-(p_y^\mu+\mey)^2<\left(250\GeV\right)^2$, 
      where the indices $x$ and $y$ refer to the two coordinates in the plane transverse to the beams.
\end{itemize}
The last requirement is designed to suppress the contribution from 
mismeasured muon momentum associated with large \met.
We also veto events with a second isolated electron or muon in the final state. 

Events with at least four jets, each with $\pt>20\GeV$
  and $|\eta|<2.5$, are accepted for further analysis. 
The leading jet, i.e., the jet of highest \pt, is required to have $\pt>40\GeV$. 
As in Ref.~\cite{bib:D0xsect}, we minimize the effect of multiple collisions in the same
bunch crossing by requiring that jets have at least two tracks
within the jet cone pointing back to the PV. 
We also require that at least one of the four leading jets is identified as a $b$ jet. 

The main background after this event selection is from the production of $W$ bosons in association with jets (\wpj).  
There is a small contribution from multijet (MJ) production, where jets are misidentified as leptons. 
Other small backgrounds from single top quark and diboson production 
are insignificant for this analysis~\cite{bib:p17PRL}.
We use the \mcatnlo\ event generator~\cite{bib:mcatnlo} combined with \herwig\ showering~\cite{bib:herwig} 
to model the behavior of \ttbar\ events, and
\alpgen~\cite{bib:alpgen} combined with \pythia~\cite{bib:pythia} 
to simulate the \wpj\ background.
The events generated by the Monte Carlo (MC) programs are
passed through the \DZ\ detector simulation~\cite{bib:run2det} and 
the same reconstruction that was used for data. 
To model energy depositions from noise and additional \ppbar\ collisions
within the same bunch crossing, data from random \ppbar\ crossings 
are overlaid over the simulated events.
The properties of the MJ background are evaluated using
control samples from \DZ\ data.

%
%
\subsection{Kinematic reconstruction}

The kinematic characteristics of each \ttbar\ event are determined 
from the decay products through a constrained kinematic fit to the \ttbar\ hypothesis~\cite{bib:hitfit}. 
In the kinematic fit, the energies and angles of the detected objects are varied and the most likely 
jet--parton assignment is identified by minimizing a \chisq\ function based on the experimental resolution. 
Since the resolution on \met\ is much worse than on any other reconstructed object, 
we do not include a constraint from \met\ in the \chisq\ calculation.
In the fit, the lepton momentum and \met,
as well as energies of two of the jets, are constrained to combine to 
objects with invariant masses of $80.4\GeV$, the mass of $W$ boson. 
Additionally, the invariant masses of the hadronic and leptonic top quark candidates, each a combination of detected objects, are constrained to be $172.5\GeV$~\cite{bib:topmass}.  
 
The four leading jets are considered in the kinematic fit. 
The $b$-tagging information is used to reduce the
number of jet assignments considered in the kinematic fit by requiring that a $b$-tagged jet can only be
assigned to $b$ quarks from top quark decay.

We retain the events in which the kinematic fit converges and further analyze the most likely 
jet--parton assignment for each event. 
The kinematic fit converges more than 99\% of the time. 
It identifies the correct assignment in $\approx 70$\% of the simulated events where each quark 
from \ttbar\ decay yields one of the jets considered in the kinematic fit. 
The distribution of the minimal \chisq\ is presented in Fig.~\ref{fig:invars}(a) and shows
good agreement between data and simulation. 

%
%
\subsection{Defining the asymmetries}

We define the difference in rapidities between 
the top quark and antitop quark,
\begin{equation}
\dy = y_t - y_\tbar = \qlep(\ytl - \yth),
\label{eq:dy}
\end{equation}
where \qlep\ is the charge of the lepton, and \ytl\ (\yth) is the rapidity of the 
leptonic (hadronic) top quark.
The corresponding forward-backward asymmetry is:
\begin{equation}
\afb = \frac {\Nf - \Nb} {\Nf + \Nb},
\label{eq:afb}
\end{equation}
where \Nf\ is the number of ``forward'' events with $\dy > 0$, 
and \Nb\ is the number of ``backward'' events with $\dy < 0$.
The rapidity difference is invariant under boosts along the beam axis,
and \afb\ corresponds to the asymmetry in the \ttbar\ rest frame. 

In addition, we consider an asymmetry based on the charge and rapidity (\ylep) of the 
electron or muon originating from the $W$ boson from top quark decay: 
\begin{equation}
\afbl = \frac {\Nfl - \Nbl} {\Nfl + \Nbl},
\label{eq:afbl}
\end{equation}
where \Nfl\ is the number of events that have $\qyl > 0$,
and \Nbl\ is the number of events with  $\qyl < 0$.

The numbers of events and the asymmetries can be defined at the ``production level'',
yielding the generated, inclusive asymmetries that are comparable to the QCD calculations.
They can also be defined after event selection and reconstruction:
we report the ``raw'' numbers of forward and backward data
events before background subtraction, 
and also the ``reconstruction level'' \ttbar\ asymmetries.

%
%
\section{The predicted standard model asymmetries}
\label{sec:Preds}

As the asymmetry first appears at order $\alpha_s^3$ of the strong coupling, with the 
largest contribution due to a loop diagram,
it is not fully simulated by tree-level event generators.
In addition, the modeling of selection and reconstruction effects requires 
that the production of all long-lived particles is fully simulated.
The \mcatnlo\ event generator is well suited for this measurement as it couples 
an NLO calculation of \ttbar\ production with 
subsequent parton showers to fully simulate \ttbar\ events.
Its predictions for the asymmetry are listed in Table~\ref{tab:preds}.

The asymmetries predicted by \mcatnlo\ are smaller at the reconstruction level due to
several effects. The event selection has a higher efficiency for events with $\dy<0$ 
than for those with $\dy>0$, lowering \afb. 
Due to the correlation between \afb\ and \afbl, the acceptance also lowers \afbl.
Finally, the limited experimental resolution on \dy\ reduces $\left|\afb\right|$.

\begin{table}[htbp]
\caption{Predictions from \mcatnlo.
  \label{tab:preds}
}
\begin{ruledtabular}
\begin{tabular}{llcc}
Level & Channel & \afb\ (\%) & \afbl\ (\%) \\
\hline \tablestrut 
Production & \lpj & $5.0\pm0.1$ & $2.1\pm0.1$ \\[2ex]
Reconstruction 
         & \epj & $2.4\pm0.7$ & $0.7\pm0.6$ \\
         & \mpj & $2.5\pm0.9$ & $1.0\pm0.8$  \\
         & \lpj & $2.4\pm0.7$ & $0.8\pm0.6$ \\
\end{tabular}
\end{ruledtabular}
\end{table}

Including the $\alpha_s^4$ terms in the calculation of \afb\ for $\ttbar j$ processes
yields an asymmetry that is significantly less negative than at order $\alpha_s^3$~\cite{bib:dittmaier}. 
Reference~\cite{bib:melnikov} argues that this does not affect the inclusive asymmetry in \ttbar\ production.
\mcatnlo\ simulates top quark decays only in LO.
Recent calculations, which include additional terms missing from
the \mcatnlo\ matrix elements and/or threshold resummations, find 
\afb\ values of $5$ to $9$\%~\cite{bib:kidonakis,bib:bernsi,bib:ahrens,bib:hollik}
and \afbl\ values of $\approx 3.5$\%~\cite{bib:bernsi}.
The uncertainties on the calculated \afb s due to the choice of renormalization and factorization
scales are below 1\%.

%
%
\section{\boldmath Measuring the reconstructed \afb}
\label{sec:fitafb}
The procedure for estimating the background and measuring \afb\ is similar 
to the one used in Ref.~\cite{bib:p17PRL}. 
To estimate the amount of background from \wpj\ production, 
we define a ``likelihood''~\cite{bib:Whel} discriminant
using variables that are modeled well by our simulation, provide separation
between signal and \wpj\ events, and do not bias \absdy\ for the signal.
The last criterion is specific to the \afb\ measurements, as many of the common
variables used to discriminate between \ttbar\ production and \wpj\ production
are biased towards central events, and therefore towards
small \absdy\ values that are less suited for this measurement.

Figure~\ref{fig:invars} shows the distributions of the four variables
chosen as inputs to the discriminant: 
(a) \chisq\ of the solution chosen by the constrained kinematic fit, 
(b) transverse momentum of the leading $b$-tagged jet, 
(c) $\ktmin=\min\left(p_T^1,p_T^2\right)\cdot\Delta R^{12}$, where
$\Delta R^{12}$ is the distance in the $\eta$-$\phi$ plane
 between the two closest jets, and $p_T^1$ and $p_T^2$ are 
their transverse momenta,
(d) the invariant mass of the jets assigned to the $W\to q \qbar'$ decay in the kinematic fit, 
calculated using kinematic quantities before the fit.
\chisq\ and \mjj\ indicate how well the event matches the
$\ttbar\to\lpj$ hypothesis.
Jets in \wpj\ and MJ background are often
due to a hard gluon emitted from a final state parton;
such jets tend to have low \ktmin\ values.
Lastly, \lbpt\ exploits the kinematic differences between
$b$ jets from top decays and those from gluon splitting
in \wpj\ and MJ events.

The amounts of \ttbar, \wpj, and MJ background shown in the figures
are taken from the fit described below.

\begin{figure*}[htbp]
\begin{center}
\includegraphics[scale=.39]{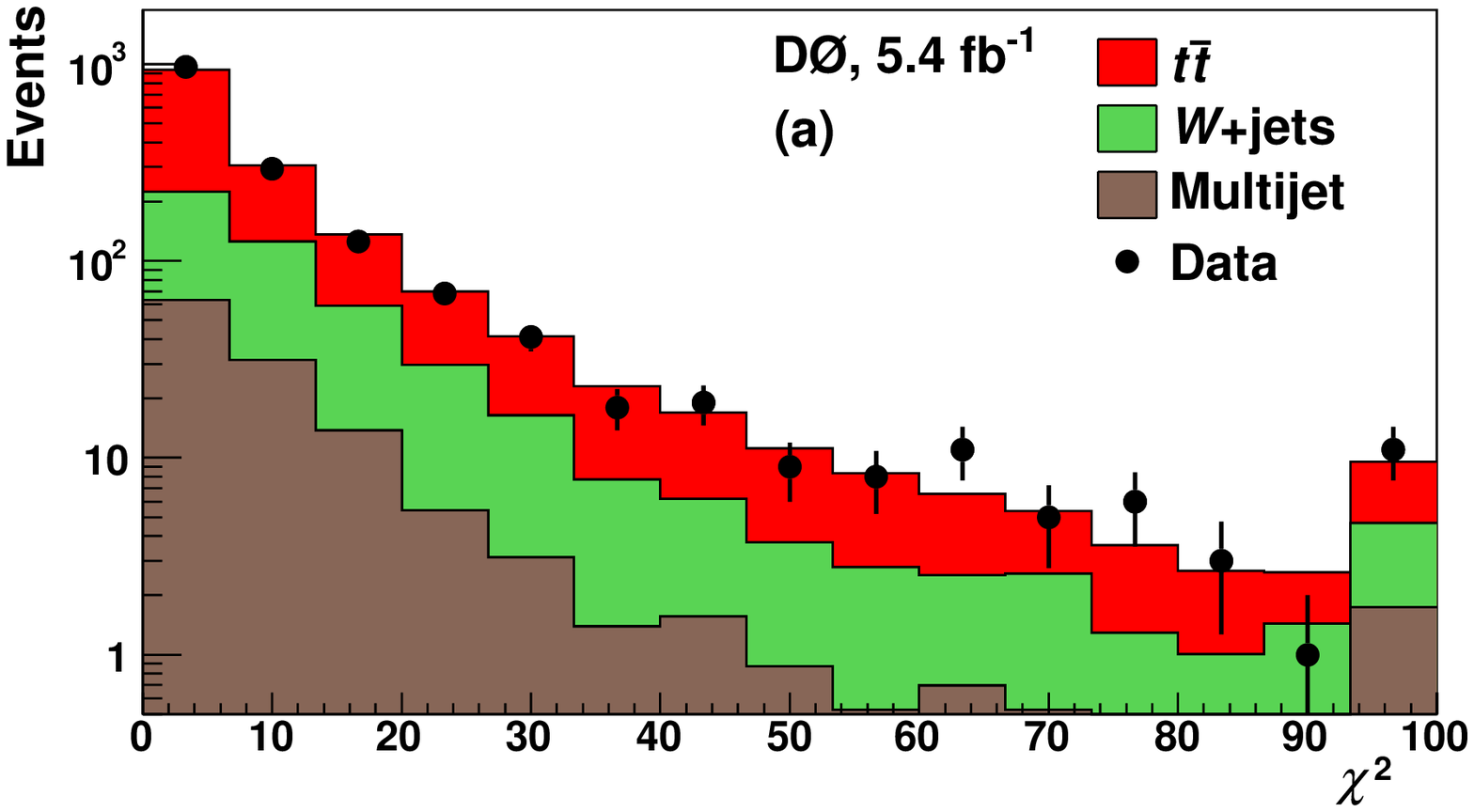}
\includegraphics[scale=.39]{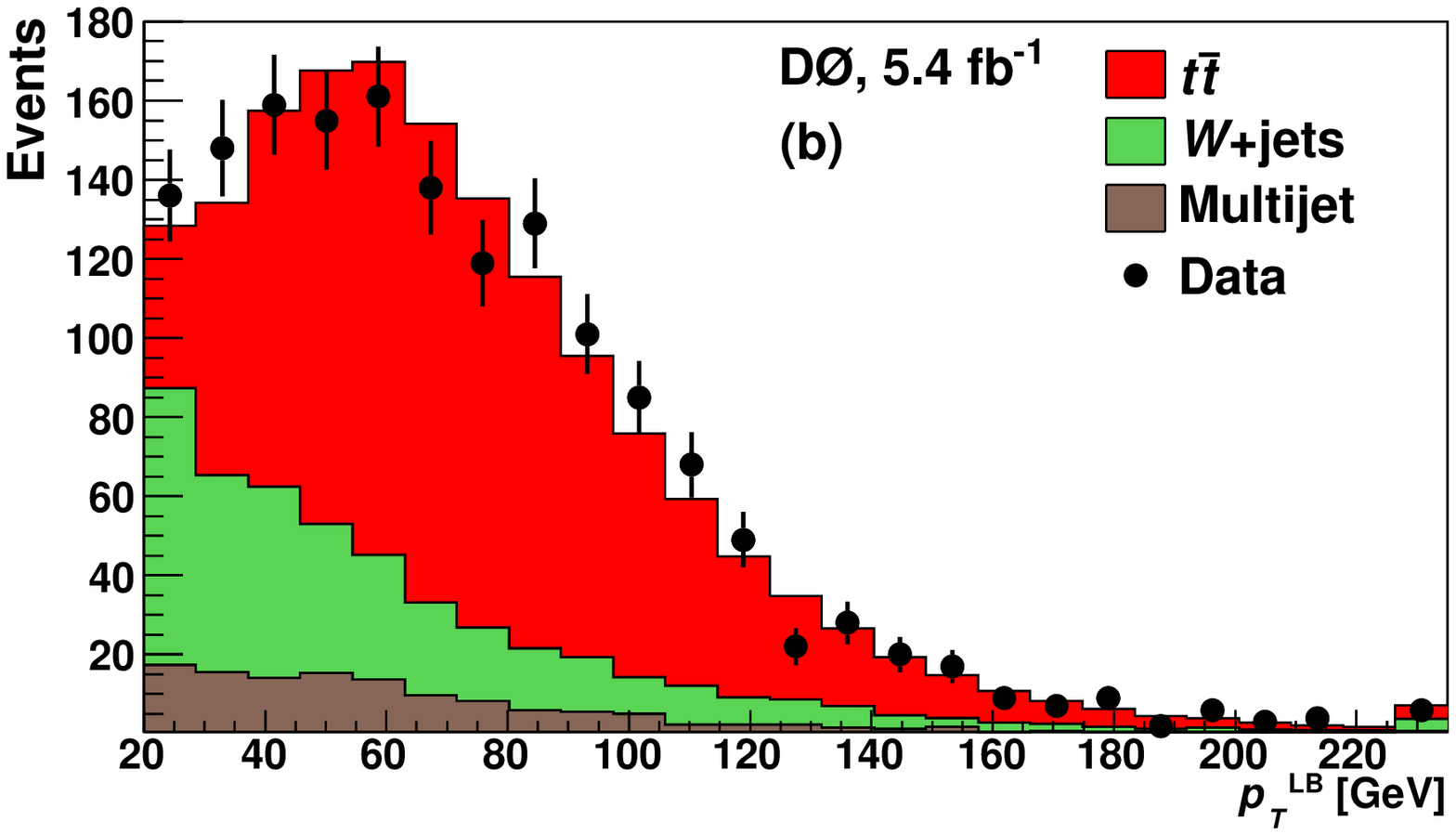}\\
\includegraphics[scale=.39]{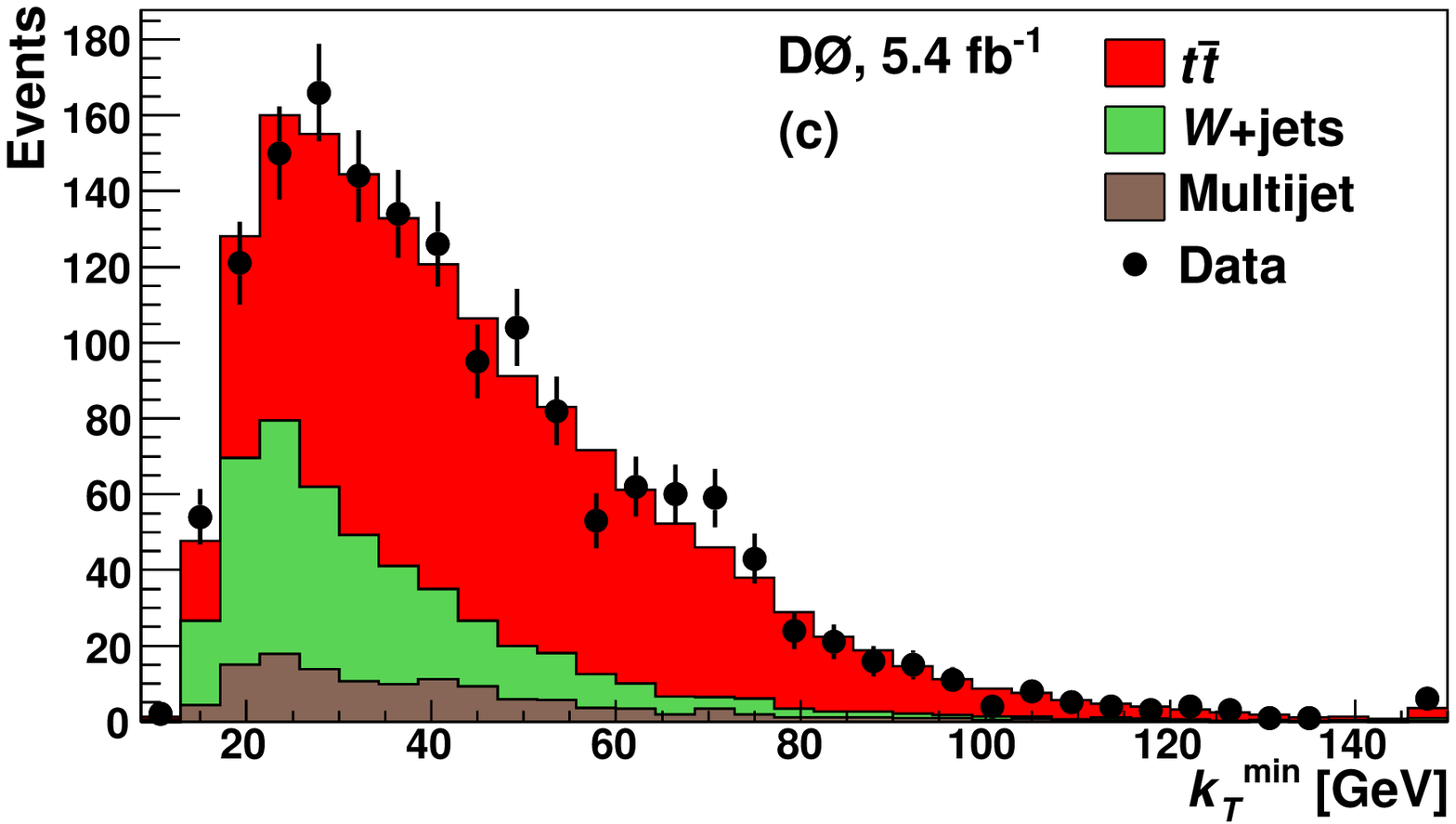}
\includegraphics[scale=.39]{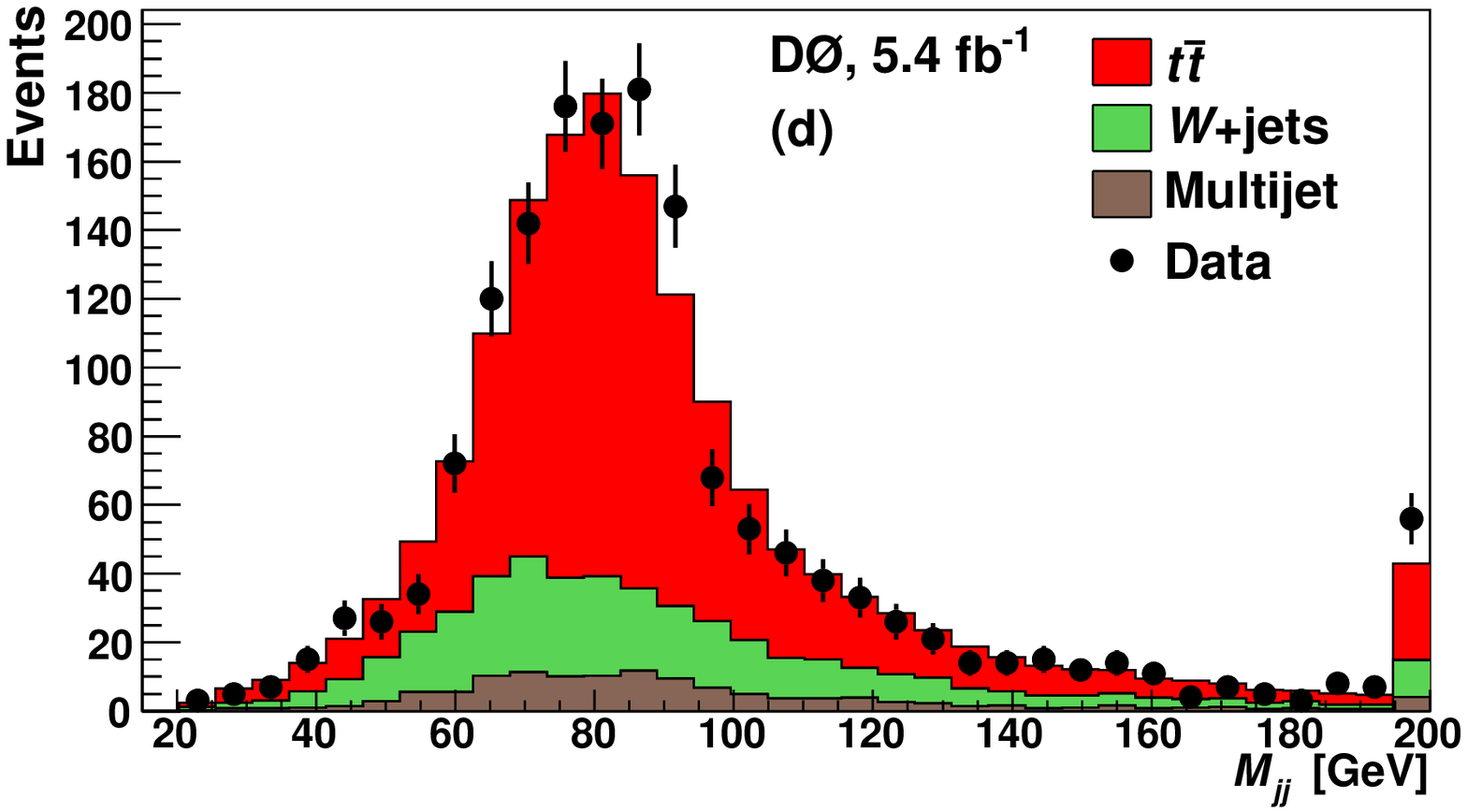}\\
\end{center}
\vspace{-0.6cm}
\caption{
Input variables to the discriminant between \ttbar\ and background events.
Overflows are added to the highest bins.
}
\label{fig:invars}
\end{figure*}

The composition of the data sample and the reconstructed \afb\ are extracted simultaneously using a 
maximum likelihood fit to the distributions of the discriminant and $\sgn(\dy)$.
The following four samples are used to construct the templates for the fit:
\begin{itemize}
\item simulated \ttbar\ events with $\dy>0$ (the $t$ quark is reconstructed as more forward than the \tbar\ quark),
\item simulated \ttbar\ events with $\dy<0$ (the \tbar\ quark is reconstructed as more forward than the $t$ quark),
\item simulated \wpj\ events,
\item a control data sample that has been enriched in MJ production
      by inverting the lepton isolation requirements~\cite{bib:D0xsect}.
\end{itemize}
The distribution of the discriminant is the same for both \ttbar\ templates.
The normalization of the MJ background is evaluated using data based on the 
probability of a jet to satisfy the lepton quality requirements~\cite{bib:D0xsect}. 
The likelihood maximized in the fit
relates the relative normalization of the first two templates to \afb,
so that the fitted \afb\ describes the reconstruction level asymmetry after
background subtraction.

Table~\ref{tab:afbcomb} summarizes the results of maximum likelihood fits to the full dataset and to several subsamples 
selected based on lepton flavor and on the number of jets in the event. 
Templates are derived separately for each subsample.  
The distributions of the discriminant are shown in Fig.~\ref{fig:disc}
and the distribution of \dy\ is shown in Fig.~\ref{fig:dy_incl}.
The fitted asymmetries are higher than predicted in all of these samples, except for the $l$+$\ge$5 jet sample.

\begin{figure}[htbp]
\begin{center}
\includegraphics[scale=.39]{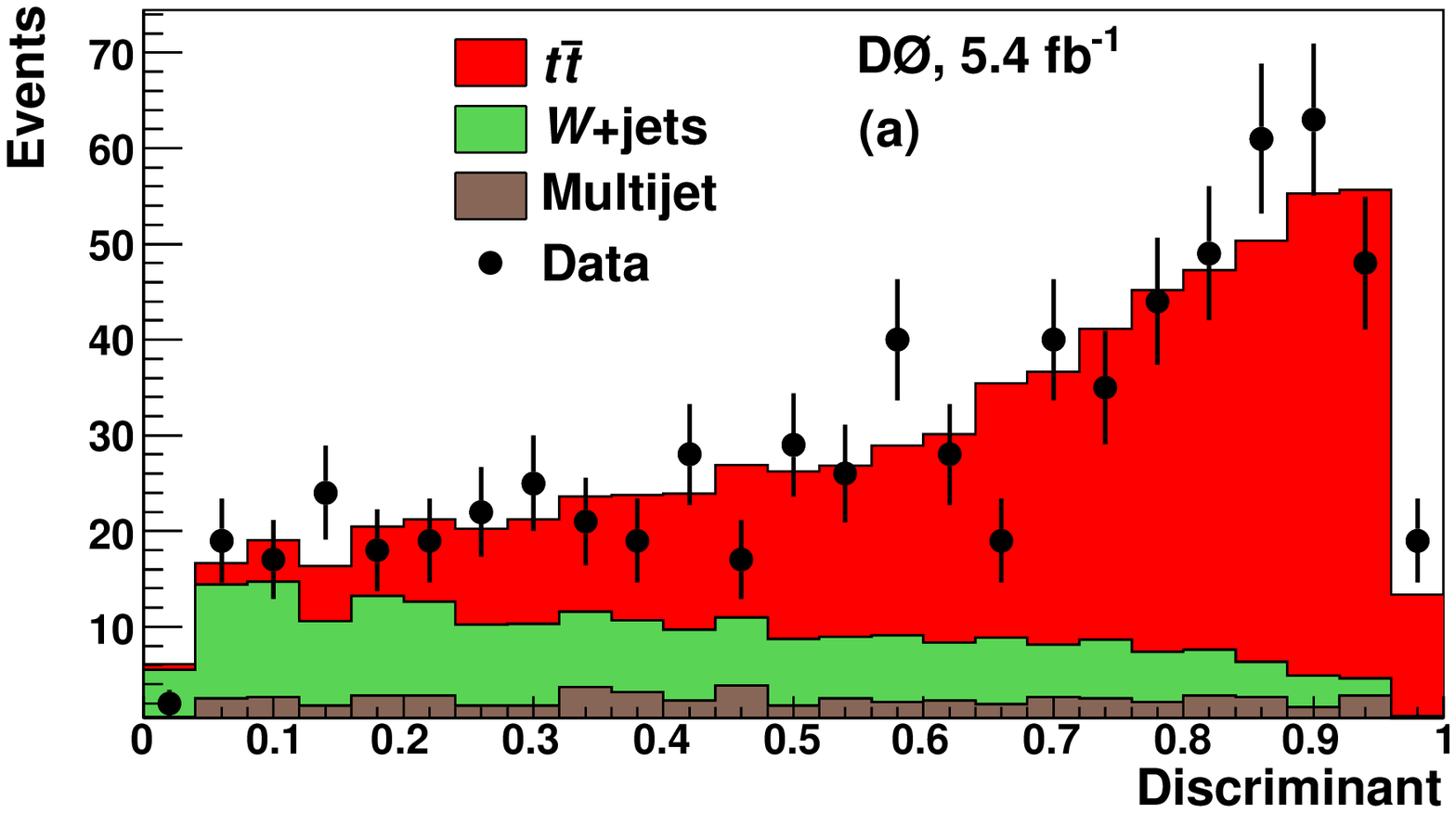}
\includegraphics[scale=.39]{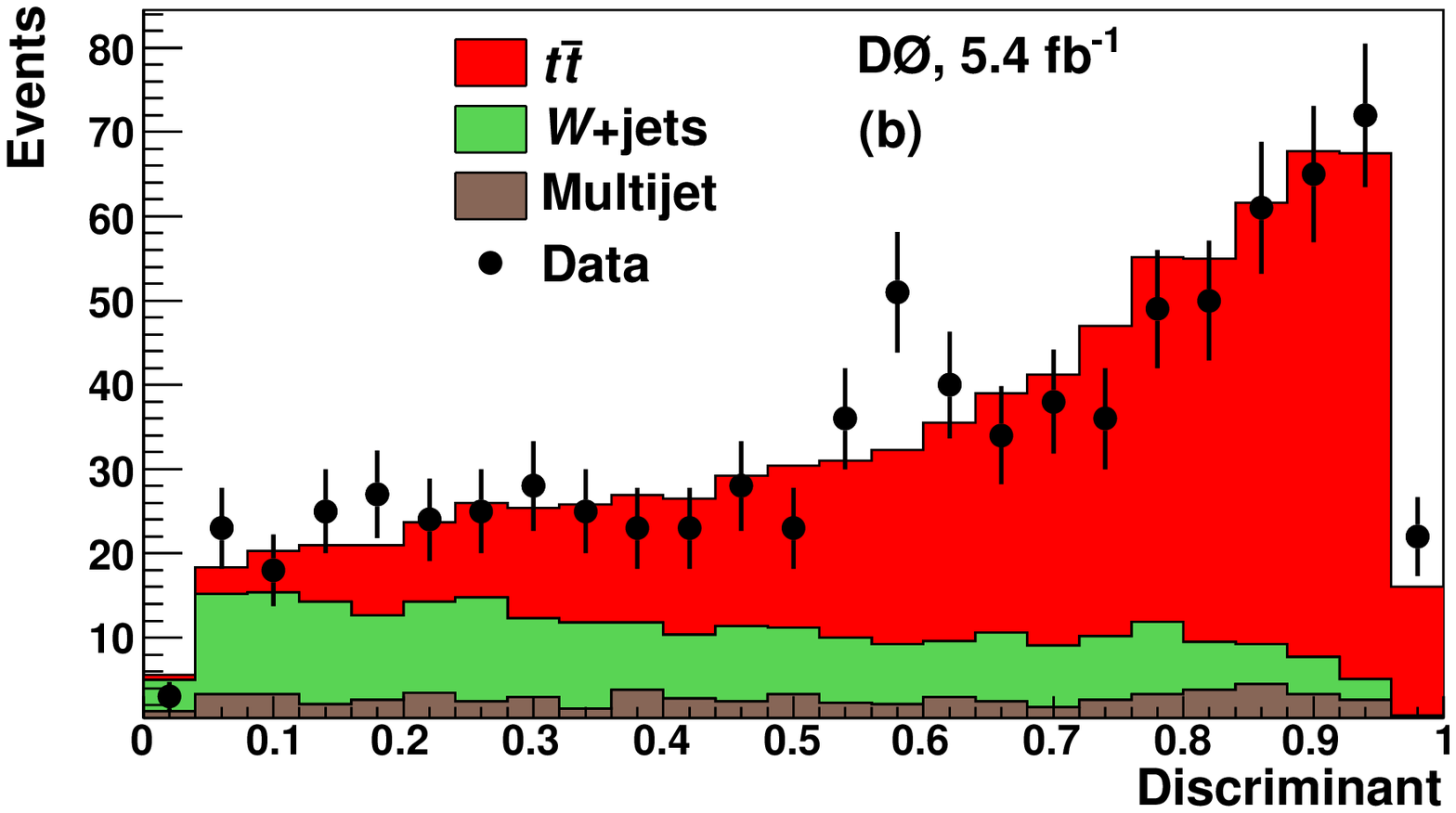}
\end{center}
\vspace{-0.6cm}
\caption{
The discriminant for events with (a) $\dy<0$ and (b) $\dy>0$.
}
\label{fig:disc}
\end{figure}

\begin{figure}[htbp]
\begin{center}
\includegraphics[scale=.39]{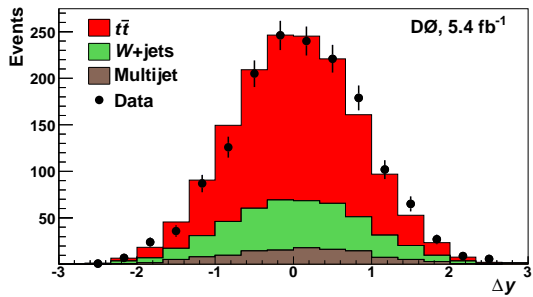}
\end{center}
\vspace{-0.6cm}
\caption{
The reconstructed \dy.
Bin widths correspond to about half of the detector resolution in \dy.
}
\label{fig:dy_incl}
\end{figure}

\begin{table*}[htbp]
\caption{Numbers of events in data, results of fits for sample composition and \afb, 
         and predictions for \afb. The asymmetries are given at reconstruction level,
         with their total uncertainties. The sample compositions are given with 
         the fit uncertainties.
  \label{tab:afbcomb}
}
\begin{ruledtabular}
  \begin{tabular}{l*{5}{D{,}{\,\pm\,}{5.5}}}
    & \multicolumn{1}{c}{$l$+$\ge$4 jets} & \multicolumn{1}{c}{$e$+$\ge$4 jets} & 
      \multicolumn{1}{c}{$\mu$+$\ge$4 jets} & \multicolumn{1}{c}{$l$+4 jets} & \multicolumn{1}{c}{$l$+$\ge$5 jets} \\
    \hline \tablestrut 
    Raw \Nfwd & \multicolumn{1}{c}{849} & \multicolumn{1}{c}{455} &
      \multicolumn{1}{c}{394} & \multicolumn{1}{c}{717} & \multicolumn{1}{c}{132}\\
    Raw \Nbwd & \multicolumn{1}{c}{732} & \multicolumn{1}{c}{397} & \multicolumn{1}{c}{335} &
      \multicolumn{1}{c}{597} & \multicolumn{1}{c}{135}\\[2ex]
    \Ntt       & 1126,39 & 622,28 & 502,28 & 902,36 & 218,16\\
    \Nw        &  376,39 & 173,28 & 219,27 & 346,36 &  35,16\\
    \Nmj       &   79,5 &  56,3 &   8,2 &  66,4 &  13,2\\
    \afb (\%) &    9.2,3.7 &   8.9,5.0 &   9.1,5.8 &  12.2,4.3 &  -3.0,7.9\\[2ex]
    \mcatnlo\ \afb\ (\%) &    2.4,0.7 &  2.4,0.7   &   2.5,0.9  &  3.9,0.8 &  -2.9,1.1\\
  \end{tabular}
\end{ruledtabular}
\end{table*}

Contributions from physics beyond the SM can modify the dependence of \afb\ on the kinematics of the \ttbar\ system. 
For example, the presence of a heavy mediator in the $s$ channel of \ttbar\ production could enhance the dependence of 
the \afb\ on the invariant mass of the \ttbar\ system (\mttbar), while contributions from $t$-channel 
production~\cite{bib:gresham} and from box diagrams could also enhance its dependence on \absdy.

The recent CDF measurement~\cite{bib:CDFdep} found
an enhanced asymmetry in regions with high \mttbar\ and in regions with large \absdy.
\mcatnlo\ predicts that the asymmetry is enhanced for high \mttbar\ and for large \absdy,
but by amounts that are small compared to the current experimental precision.
In Table~\ref{tab:subsamples} we summarize our measurement of these 
dependencies. We do not find any statistically significant dependencies.

\begin{table}[htbp]
\caption{
  Reconstruction-level \afb\ by subsample.
  \label{tab:subsamples}
}
\begin{ruledtabular}
\begin{tabular}{l*{2}{D{,}{\,\pm\,}{-1}}}
  &  \multicolumn{2}{c}{\afb\ (\%)} \\
Subsample  & \multicolumn{1}{c}{\hspace{1ex}Data} & \multicolumn{1}{c}{\hspace{1em}\mcatnlo} \\ 
\hline \tablestrut 
$\mttbar<450\GeV$ &  7.8 , 4.8 & 1.3 , 0.6  \\   
$\mttbar>450\GeV$ & 11.5 , 6.0 & 4.3 , 1.3  \\
$\absdy<1.0$      &  6.1 , 4.1 & 1.4 , 0.6  \\
$\absdy>1.0$      & 21.3 , 9.7 & 6.3 , 1.6 \\
\end{tabular}
\end{ruledtabular}
\end{table}

%
%
\section{\boldmath Measuring the production \afb}
\label{sec:unfold}
In the previous section we discussed the measurement 
of the \ttbar\ asymmetry at the reconstruction level. 
This quantity is necessarily detector specific, which makes the interpretation of the 
result as well as comparison to theory and
to other experiments problematic. It is therefore desirable to infer the asymmetry at the 
production level by correcting for (``unfolding'') 
the effects of detector resolution and acceptance on the observed asymmetry.

Only the numbers of events produced with positive and negative \dy\ are relevant 
for the calculation of the asymmetry. The migration of events within these categories, 
due to the finite experimental resolution in \dy, does not affect the reconstructed asymmetry.
Thus, to present the result in terms of the \ttbar\ production asymmetry requires an accurate
correction of the migration across the boundary ($\dy =0$).
The importance of this correction grows with the fraction of events that fall 
within the detector resolution of the boundary.
For \dy\ this fraction is $\approx 20$\%, to be contrasted with $\approx 10$\% for
the rapidity of the hadronic top quark as used in Ref.~\cite{bib:CDFdep}, 
and with $\approx 0.1$\% for the lepton rapidity.
 
We first bin the distributions of \dy\ at the production and reconstruction levels.
The migrations from one \dy\ bin to another are described through a two-dimensional matrix,
and the acceptance through a diagonal matrix.
To accurately describe the migration between events with positive and negative \dy, 
it is desirable to have fine binning in the region where the probability to misreconstruct the sign of 
\dy\ changes rapidly, that is, near $\dy=0$~\cite{bib:p17PRL}.
Fine binning is less important at large \absdy. 
Coincidentally, the large \absdy\ region has lower statistics both in data and simulation,
thereby limiting the precision of the migration matrix, which is derived from simulated events. 
To reduce this effect, we use bins of variable size, increasing towards large \absdy. 
We bin the \dy\ distribution in 50 bins at the reconstruction level and 
in 26 bins at the production level. 

In general, unfolding histograms where the bin width is smaller than the experimental
resolution is unstable with respect to statistical fluctuations in the data.
Regularization techniques are employed to suppress such fluctuations by
smoothing the unfolded results~\cite{bib:zechbook}.

We find the generated \dy\ distribution using a regularized unfolding,
and then summarize this distribution into the \afb\ observable according to Eq.~\ref{eq:afb}.
The unfolding is implemented using the \tunfold\ software~\cite{bib:tunfold},
which we modified to account for variable bin widths.

In References~\cite{bib:CDFPRL, bib:CDFdep} the need for an explicit regularization 
is avoided by using wide bins in \dy\ with boundaries at $\dy=-3$, $-1$, 0, 1, and 3. 
The unfolding then reduces to inverting a 4-by-4 matrix.
This implicit regularization averages out migrations (and acceptance) in the 
wide \dy\ range of each bin, with the disadvantage that the migration across the $\dy=0$ boundary
is under-estimated for events near $\dy=0$ while it is over-estimated for events
near the outer edges of the central bins. 

Since the regularization suppresses the badly-measured components of the data, it can
also suppress part of the \ttbar\ production asymmetry.
We calibrate the unfolding using ensembles of pseudo-datasets (PDSs). 
Each PDS is generated including signal and background contributions
and is unfolded using the same procedure as for \DZ\ data.
We use the \dy\ distribution of \ttbar\ events predicted by \mcatnlo\ and a wide
variety of distributions inspired by the scenarios beyond the SM, which were listed in the introduction. 
We choose a regularization strength that balances the statistical strength of the measurement 
and its model dependence. 
We find that the unfolded asymmetries are smaller than the input values by
a multiplicative factor of $0.93\pm0.05$, where the uncertainty 
covers the various scenarios with $\afb>5$\% and the SM scenario.
All values and uncertainties given for the unfolded \afb\ are corrected for this bias,
and the uncertainty in this factor is propagated to the result.

We estimate the statistical uncertainty on the unfolded asymmetry from its
RMS in an ensemble based on the \mcatnlo\ prediction. 
The regularized fine-bin unfolding results in a statistical uncertainty 
on \afb\ of 6.0\%, while the coarse-bin matrix inversion technique~\cite{bib:CDFPRL, bib:CDFdep}
results in a statistical uncertainty of 7.7\%. 
The results of the fine-bin unfolding are given in Table~\ref{tab:dyafb}. 
For comparison, the 4-bin unfolding procedure yields $\afb=\left(16.9\pm 8.1\right)$\%,
with the statistical and systematic uncertainties combined.

\begin{table}[htbp]
\caption{
  \dy-based asymmetries. 
  \label{tab:dyafb}
}
\begin{ruledtabular}
\begin{tabular}{lD{,}{\,\pm\,}{7.7}D{,}{\,\pm\,}{7.5}}
  &  \multicolumn{2}{c}{\afb\ (\%)} \\
  & \multicolumn{1}{c}{\hspace{-1ex}Reconstruction level} & \multicolumn{1}{c}{Production level} \\ 
\hline \tablestrut 
Data     & 9.2,3.7 & 19.6,6.5 \\
\mcatnlo & 2.4,0.7 &  5.0,0.1 \\
\end{tabular}
\end{ruledtabular}
\end{table}

The difference between measured and predicted asymmetries at the production level has a 
statistical significance that corresponds to 2.4 SD, 
while it is 1.9 SD at the reconstruction level. 
Given the SM hypothesis, the probability to have this or a larger difference in significance 
between the reconstruction and production levels is 43\%.  

%
%
\section{Measuring the Lepton-based Asymmetry}
\label{sec:afbl}
An alternative to measuring and unfolding \afb\ is to
measure the asymmetry \afbl, defined in Eq.~\ref{eq:afbl}.
The procedure to measure \afbl\ at the reconstruction level is identical to that for \afb.
Figure~\ref{fig:ylq} shows the distribution of \qyl.
In simulated \ttbar\ events, the correlation between \qyl\ and the reconstructed \dy\ is $38$\%.
Background subtraction is performed using a fit for events selected with an additional requirement of $\absyl<1.5$, as described below. 
The results of the fit are given in Table~\ref{tab:afbl}.
\begin{figure}[htbp]
\begin{center}
\includegraphics[scale=.39]{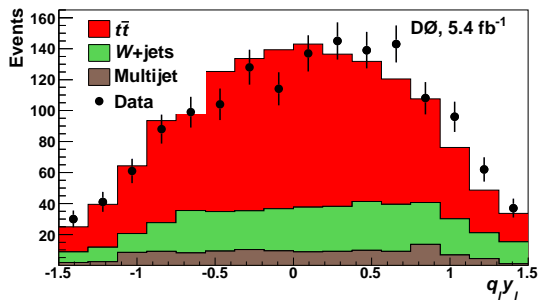}
\end{center}
\vspace{-0.6cm}
\caption{
The reconstructed charge-signed lepton rapidity.
}
\label{fig:ylq}
\end{figure}

\begin{table*}[htbp]
\caption{Numbers of events in data, results of fits for sample composition and \afbl, 
         and predictions for \afbl. The asymmetries are given at reconstruction level,
         with their total uncertainties. The sample compositions are given with 
         the fit uncertainties.
  \label{tab:afbl}
}
\begin{ruledtabular}
  \begin{tabular}{l*{5}{D{,}{\,\pm\,}{5.5}}}
    & \multicolumn{1}{c}{$l$+$\ge$4 jets} & \multicolumn{1}{c}{$e$+$\ge$4 jets} & 
      \multicolumn{1}{c}{$\mu$+$\ge$4 jets} & \multicolumn{1}{c}{$l$+4 jets} & \multicolumn{1}{c}{$l$+$\ge$5 jets} \\
    \hline \tablestrut 
    Raw \Nlfwd & \multicolumn{1}{c}{867} & \multicolumn{1}{c}{485} & 
      \multicolumn{1}{c}{382} & \multicolumn{1}{c}{730} & \multicolumn{1}{c}{137}\\
    Raw \Nlbwd & \multicolumn{1}{c}{665} & \multicolumn{1}{c}{367} & 
      \multicolumn{1}{c}{298} & \multicolumn{1}{c}{546} & \multicolumn{1}{c}{119} \\[2ex]
    \Ntt        & 1096,39 & 622,28 & 474,27 & 881,36 & 211,16\\
    \Nw         &  356,39 & 173,28 & 198,27 & 323,36 &  31,16\\
    \Nmj        &   79, 5 &  56, 3 &   8, 2 &  66, 4 &  14, 2\\
    \afbl\ (\%) &   14.2, 3.8 &  16.5, 4.9 &  9.8, 5.9 &  15.9, 4.3 &   7.0, 8.0\\[2ex]
    \mcatnlo\ \afbl\ (\%) &   0.8, 0.6 &  0.7, 0.6 &  1.0, 0.8 &  2.1,0.6 &   -3.8, 1.2\\
  \end{tabular}
\end{ruledtabular}
\end{table*}

Lepton reconstruction offers excellent angular resolution 
and accurate determination of electric charge, 
making migrations and their corrections negligible. 
By measuring this observable we therefore avoid
the complications encountered in measuring \afb, due to significant migration in \dy.

\begin{table}[htbp]
\caption{Lepton-based asymmetries. 
  \label{tab:lafb}
}
\begin{ruledtabular}
\begin{tabular}{lD{,}{\,\pm\,}{4.4}D{,}{\,\pm\,}{6.4}}
  &  \multicolumn{2}{c}{\afbl\ (\%)} \\
  & \multicolumn{1}{c}{\hspace{-1ex}Reconstruction level} & \multicolumn{1}{c}{Production level} \\ 
\hline \tablestrut 
Data     & 14.2 , 3.8 & 15.2 , 4.0 \\
\mcatnlo & 0.8 , 0.6&  2.1 , 0.1 \\
\end{tabular}
\end{ruledtabular}
\end{table}

Correcting for detector effects thus reduces to weighting each \qyl\ bin by an acceptance factor, 
which is the inverse of the selection probability.
Acceptance drops rapidly for $1.1<\absyl<2$,
where coverage is available only in the muon channel.
To avoid a large spread in the weights, which
would increase the statistical uncertainty, we measure \afbl\
using only events with $\absyl<1.5$.

We correct for acceptance in 48 equally-spaced bins,
and the results are presented in Table~\ref{tab:lafb}.
As in the previous section, statistical uncertainties are obtained 
from ensembles generated according to \mcatnlo\ predictions.

%
%
\section {Systematic Uncertainties}
\label{sec:syst}

We consider multiple sources of systematic uncertainty. 
We vary the modeling according to the evaluated uncertainty on each source
and then propagate the effect to the final result. 
Systematic uncertainties from different sources are added in quadrature to yield the 
total systematic uncertainties.
In Tables~\ref{tab:sysAfb} and~\ref{tab:sysAfbl} we list the systematic uncertainties  
in the following categories:
\begin{description}
\item[Jet reconstruction (reco)] This includes the jet reconstruction and identification efficiencies, as well
the efficiency of the two tracks requirement described in Sec.~\ref{sec:selection}. 
We also include the effect of the multiple \ppbar\ collisions within the same
bunch crossing that can yield additional jets. 
The efficiencies in simulation are set equal to those measured in data using a dijet sample.

\item[Jet energy measurement]
The jet energy scale (JES) is measured using dijet and photon+jet samples~\cite{bib:jes}.
The simulated jet energy resolution (JER) is calibrated using $Z+$jet data.

\item[Signal modeling]
Modeling of gluon radiation and color reconnection can affect the dependence of the asymmetry 
on the transverse momentum of the \ttbar\ system (\ttpt), as the extra radiation can differ 
between forward and backward events. 
This can affect the measured asymmetry through the sensitivity of the acceptance to \ttpt.  
\mcatnlo\ predicts that \afb\ depends on \ttpt, and to evaluate this uncertainty, 
we consider the possibility that \afb\ does not depend on \ttpt.
The effects of the finite Monte Carlo statistics and of the modeling of the detector 
are also taken into account.

\item[\boldmath $b$ tagging]
The $b$-tagging efficiency and mis-tagging probability, which are determined from data, affect
both the overall selection efficiency and how often the correct jet assignment is found in the
kinematic fit.

\item[Charge identification (ID)] 
The simulated rate of misidentification of lepton charge is calibrated
using $Z\to ll$ samples of same and opposite charge leptons.

\item[Background (Bg) subtraction]
The amounts of $W\!+$\linebreak jets and MJ background to be subtracted are changed within their fitted uncertainties. 
Uncertainties on the normalization of the MJ background also arise from 
the uncertainties on the lepton selection rates, which are used to evaluate the MJ background.
The rate of inclusive $W\!c\bar{c}$ and $Wb\bar{b}$ production predicted by \alpgen\ must be scaled up by a factor of $1.47$ 
to match the \leptonpj\ data~\cite{bib:D0xsect}. The uncertainty on this scale factor is estimated to be $15$\%.
The effects of the finite Monte Carlo statistics are also taken into account.

\item[Unfolding bias] As described in Sec.~\ref{sec:unfold}.

\end{description}

\begin{table}[htbp]
\caption{
  Systematic uncertainties on \afb. 
  \label{tab:sysAfb}
}
\begin{ruledtabular}
\begin{tabular}{lccc}
       &  \multicolumn{3}{c}{Absolute uncertainty~\footnote{Only uncertainties above $0.1$\% are listed.} (\%)} \\
       &  \multicolumn{2}{c}{Reconstruction level} & Prod. level \\
Source & Prediction & Measurement & Measurement\\
\hline \tablestrut 
Jet reco   &    $\pm 0.3$        &       $\pm 0.5$         &    $\pm 1.0$      \\
JES/JER   & $+0.5$            &     $-0.5$          &     $-1.3$     \\
Signal modeling   &  $\pm 0.3$          &     $\pm 0.5$            &  ${+0.3}/{-1.6}$        \\
$b$ tagging   & \emph{-}            &    $\pm 0.1$       &   $\pm 0.1$   \\
Charge ID  &    \emph{-}  & $+0.1$ & ${+0.2}/{-0.1}$     \\
Bg subtraction &  \emph{-} & $\pm0.1$ & ${+0.8}/{-0.7}$ \\
Unfolding Bias   &   \emph{-}            &  \emph{-}    & ${+1.1}/{-1.0}$\\ 
\hline
Total                            & ${+0.7}/{-0.5}$ & ${+0.8}/{-0.9}$ & $+1.8/{-2.6}$\\
\end{tabular}
\end{ruledtabular}
\end{table}

\begin{table}[htbp]
\caption{
  Systematic uncertainties on \afbl.
  \label{tab:sysAfbl}
}
\begin{ruledtabular}
\begin{tabular}{lccc}
       &  \multicolumn{3}{c}{Absolute uncertainty~\footnote{Only uncertainties above $0.1$\% are listed.} (\%)} \\
       &  \multicolumn{2}{c}{Reconstruction level} & Prod. level \\
Source & Prediction & Measurement & Measurement\\
\hline \tablestrut 
Jet reco &    $\pm 0.3$        &       $\pm 0.1$         &    $\pm 0.8$      \\
JES/JER  & $+0.1$            &     $-0.4$          &     ${+0.1}/{-0.6}$     \\
Signal modeling   &  $\pm 0.3$          &     $\pm 0.5$            &  ${+0.2}/{-0.6}$        \\
$b$ tagging   &     \emph{-}        &     $\pm 0.1$     &   $\pm 0.1$   \\
Charge ID  &    \emph{-}  & $+0.1$ & ${+0.2}/{-0.0}$     \\
Bg subtraction &  \emph{-} &  $\pm 0.3$ & $\pm 0.6$ \\
\hline
Total                            & $\pm 0.5$ & $\pm0.7$ & ${+1.0}/{-1.3}$\\
\end{tabular}
\end{ruledtabular}
\end{table}

%
%
\section{Cross checks}
\subsection {\boldmath Checks of the asymmetries simulated for $W\!+$jets background}
\label{sec:wpj}
The measured \ttbar\ asymmetries depend on the input asymmetries  
of the \wpj\ background, which are taken from the simulation. 
The production of \itW\ bosons is asymmetric and strongly correlated with \qyl.
However, the correlation with \dy\ is weaker, as \dy\ is reconstructed under the \ttbar\ hypothesis.
As this hypothesis does not match the \wpj\ events, the production asymmetry is reduced.
To study the \wpj\ background, we use events with no $b$-tagged jets, which are enriched in
\wpj\ production. Their simulation matches data well, as shown in Fig.~\ref{fig:nobtag}.

\begin{figure}[htbp]
\begin{center}
\includegraphics[scale=.39]{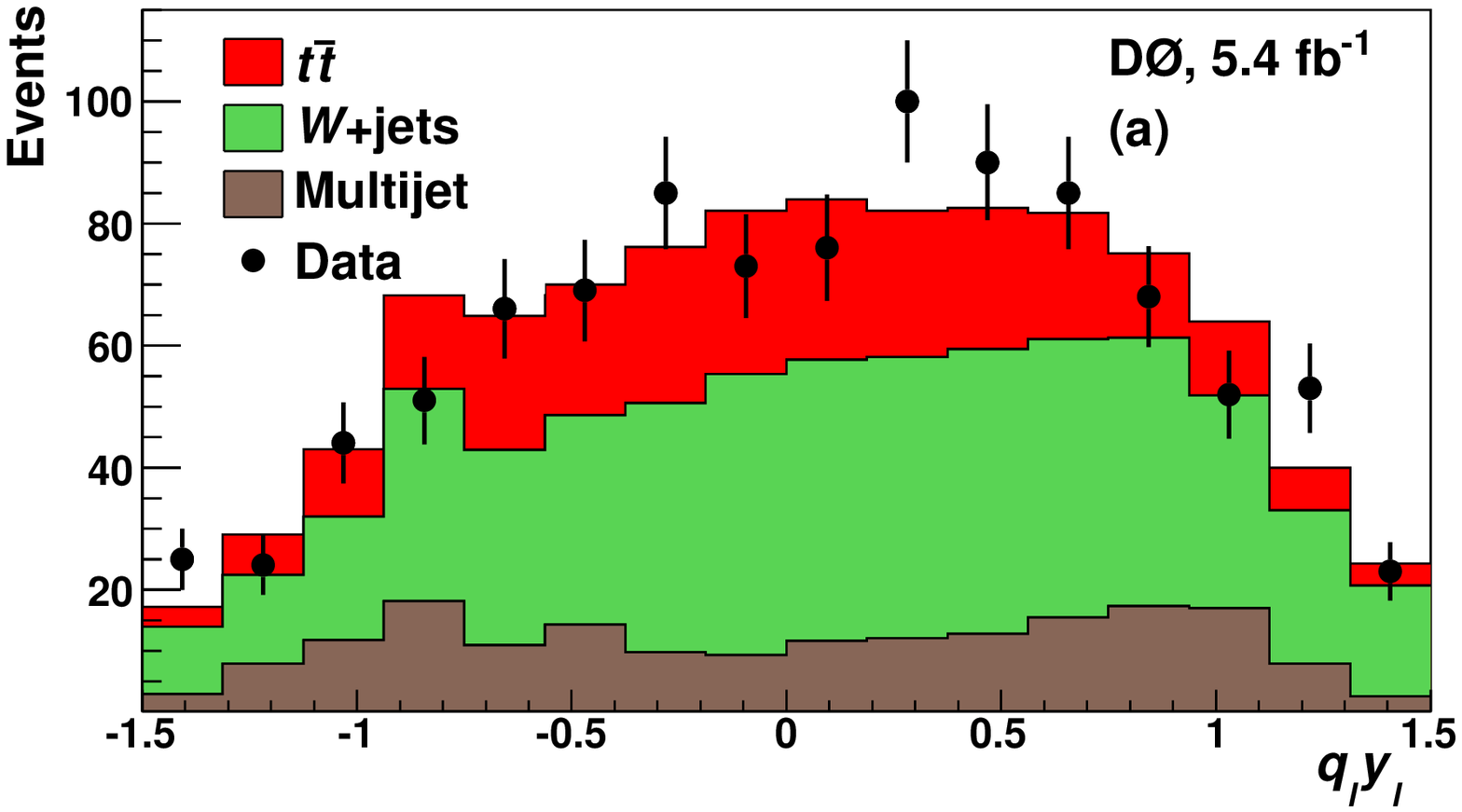}
\includegraphics[scale=.39]{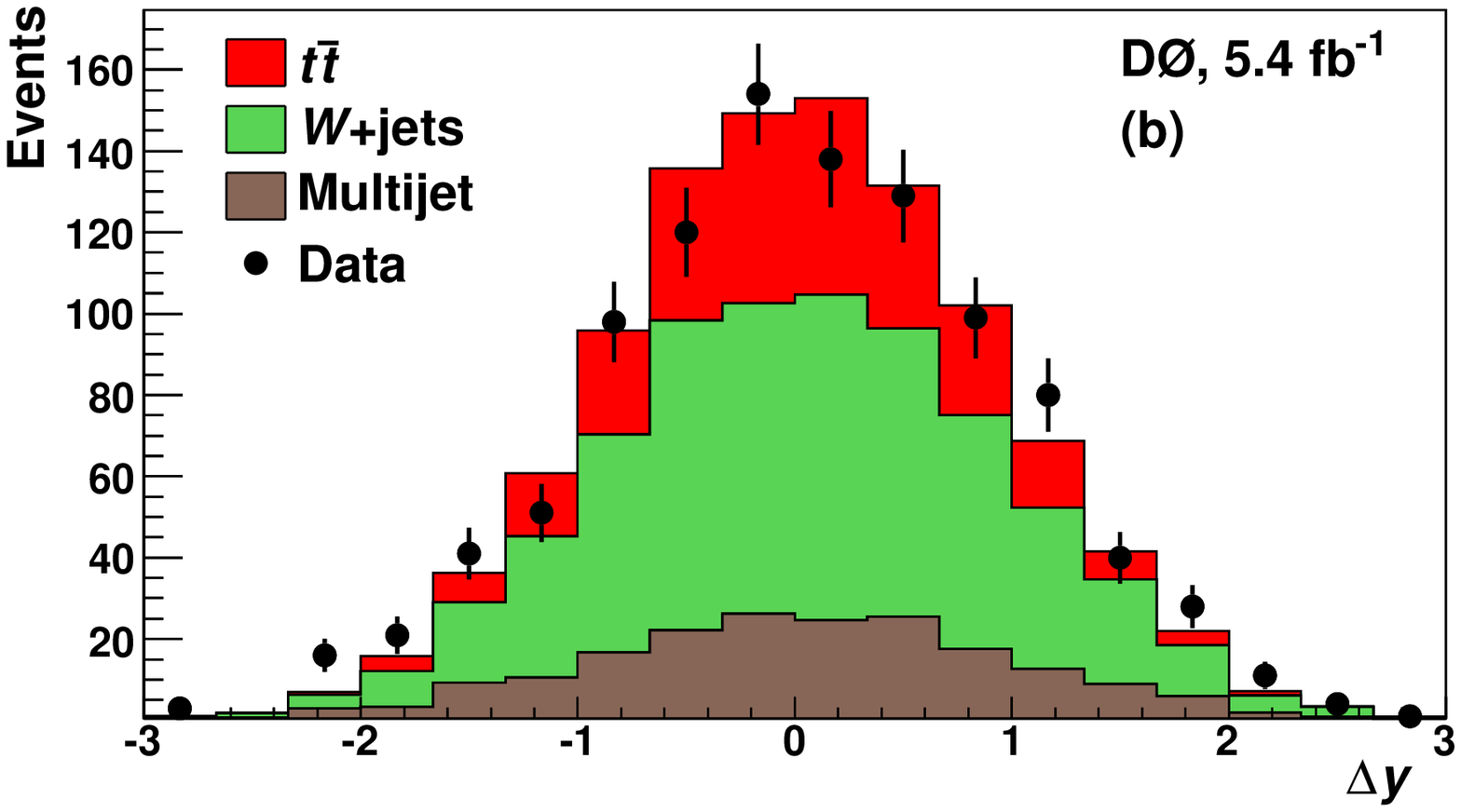}
\end{center}
\vspace{-0.6cm}
\caption{The reconstructed (a) \qyl\ and (b) \dy\ in events with no $b$-tagged jets.
In (b), the bin widths correspond to about half of the detector resolution in \dy. 
}
\label{fig:nobtag}
\end{figure}

To confirm that the asymmetries reconstructed for \wpj\ events
are properly simulated, we measure these asymmetries in data as follows.
Instead of selecting only events with at least one $b$-tagged jet,
we also select events without a $b$-tagged jet, and divide the selected events into those with $0$, $1$, 
and $\ge 2$ $b$-tagged jets.
We then perform a simultaneous fit to these samples, with the asymmetry in
the \wpj\ background as an additional fit parameter.
Some of the observables used in the fit are defined assuming that there is a $b$-tagged jet.
For the $0$-tag sample those are calculated by treating the leading jet as though it were $b$ tagged.
The fitted \wpj\ asymmetries, $\afb=\left(4.1\pm4.1\right)$\% and $\afbl=\left(15.1\pm4.1\right)$\%,
are in agreement with the simulated values of $\afb=\left(1.8\pm1.4\right)$\% and $\afbl=\left(14.3\pm1.4\right)$\%
(all uncertainties are statistical).

%
%
\subsection {Dependence on magnet polarities}
\label{sec:polar}

The polarities of the \DZ\ magnets, both the solenoid and toroid, are regularly and independently switched 
to minimize the potential impact of differences in detector acceptance and efficiency for 
positive and negative particles.
With fixed magnet polarities, localized detector problems may produce a bias, especially for \afbl.
We find no significant differences between \afbl\ values measured in subsamples with different solenoid and toroid polarities.

%
%
\subsection {Dependence on lepton charge}
\label{sec:sub_lep}
We measure \afbl, at reconstruction level, separately for events with positive 
and negative lepton charge. We find \afbl\ = $\left(12.7 \pm 5.5\right)$\% for events 
where the lepton charge is positive and \afbl\ = $\left(15.6 \pm 5.0\right)$\% 
for events where the lepton charge is negative (all uncertainties are statistical).

%
%
\section{Results and Discussion}
Tables~\ref{tab:dyafb} and~\ref{tab:lafb} summarize our measurements of the \dy- and lepton-based asymmetries 
at the reconstruction and production levels. 
The measurements are significantly higher than the \mcatnlo-based predictions.

Within the SM, the \ttbar\ production asymmetry first arises at order $\alpha_s^3$ 
as a result of interference of several production diagrams.
At this order, interference of the Born and box diagrams results in positive asymmetry in two-body production, 
while negative contributions to the asymmetry arise from $\ttbar g$ production
with a hard gluon ($\ttbar g$ production with a soft gluon is included with
the two-body production process to cancel the infrared divergence).  
Thus, the asymmetry is likely to show 
a dependence on variables that indicate the presence of extra gluons, in particular 
the multiplicity and kinematics of additional jets. As shown in Table~\ref{tab:afbcomb}, 
the asymmetry in the lepton+4 jets subsample is observed to be positive, 
while its most likely value is negative in the lepton+$\ge$5 jets subsample.

An extra parton does not always result in the reconstruction of an extra jet, 
which is required to exceed a prescribed energy threshold, and be within the acceptance of the detector. 
In particular, a gluon emitted by an initial state parton is likely to be too forward 
and/or too soft to be registered as a jet. 
The transverse momentum of the \ttbar\ system, on the other hand, is sensitive to both soft and hard gluon radiation. 
Low values of \ttpt\ correspond predominantly to two-body production,
while regions of large \ttpt\ correspond to three-body diagrams, which do not necessarily produce an extra reconstructed jet.
The dependence of the asymmetry on the presence of an extra jet
has been studied in the literature~\cite{bib:asymtev}, but we  
are not aware of previous studies of a dependence on \ttpt.

As shown in Fig.~\ref{fig:ttptdep}, some event generators predict that the \ttbar\ production asymmetry has a
strong dependence on \ttpt, while others do not. Even though \pythia\ is a tree level Monte Carlo generator, and thus 
cannot be used to predict the overall asymmetry in \ttbar\ production, 
we use it to study the interplay between \afb\ and \ttpt.
We found that this dependence is present in the \pythia\ tunes that force an angular coherence
between the top quarks and the initial state parton showers through the MSTP(67) parameter. 
We account for this possible dependence in the systematic 
uncertainties on the measured asymmetries due to signal modeling.

\begin{figure}[htbp]
\begin{center}
\includegraphics[scale=.39]{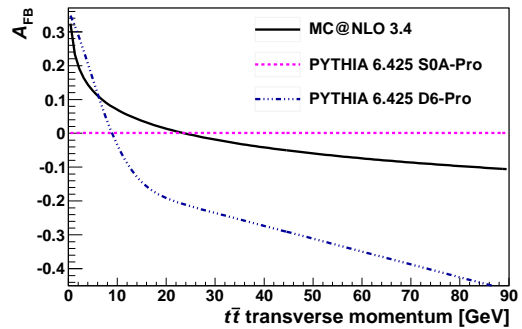}
\caption{\label{fig:ttptdep} The \ttbar\ asymmetry versus \ttpt\ as predicted by \mcatnlo+\herwig. 
For comparison, the predictions from \pythia\ with different tunes~\cite{bib:pythiatunes} are also shown.
}
\end{center}
\end{figure}

While the measured asymmetries are sensitive mostly to the well-modeled 
additional jets, we also examined the modeling of gluon radiation
with the \ttpt\ observable. No aspect of this analysis has been
optimized for this observable, and its experimental resolution
is low. Nevertheless, we note that the \ttpt\ spectrum 
is softer in data than in the \mcatnlo-based model, indicating less gluon emission, 
as shown in Figure~\ref{fig:ttpt}(a). 
To verify this hypothesis,
we simulate \ttbar\ events using \pythia\ with initial state radiation (ISR) turned off. 
In this unrealistic scenario, the \ttpt\ distribution is in better agreement with the data, as seen in Figure~\ref{fig:ttpt}(b),
but the simulated number of additional jets is too low.
In the SM, low \ttpt\ is associated with high \afb, so the two
discrepancies are in the same direction.

\begin{figure}[htbp]
\begin{center}
\includegraphics[scale=.39]{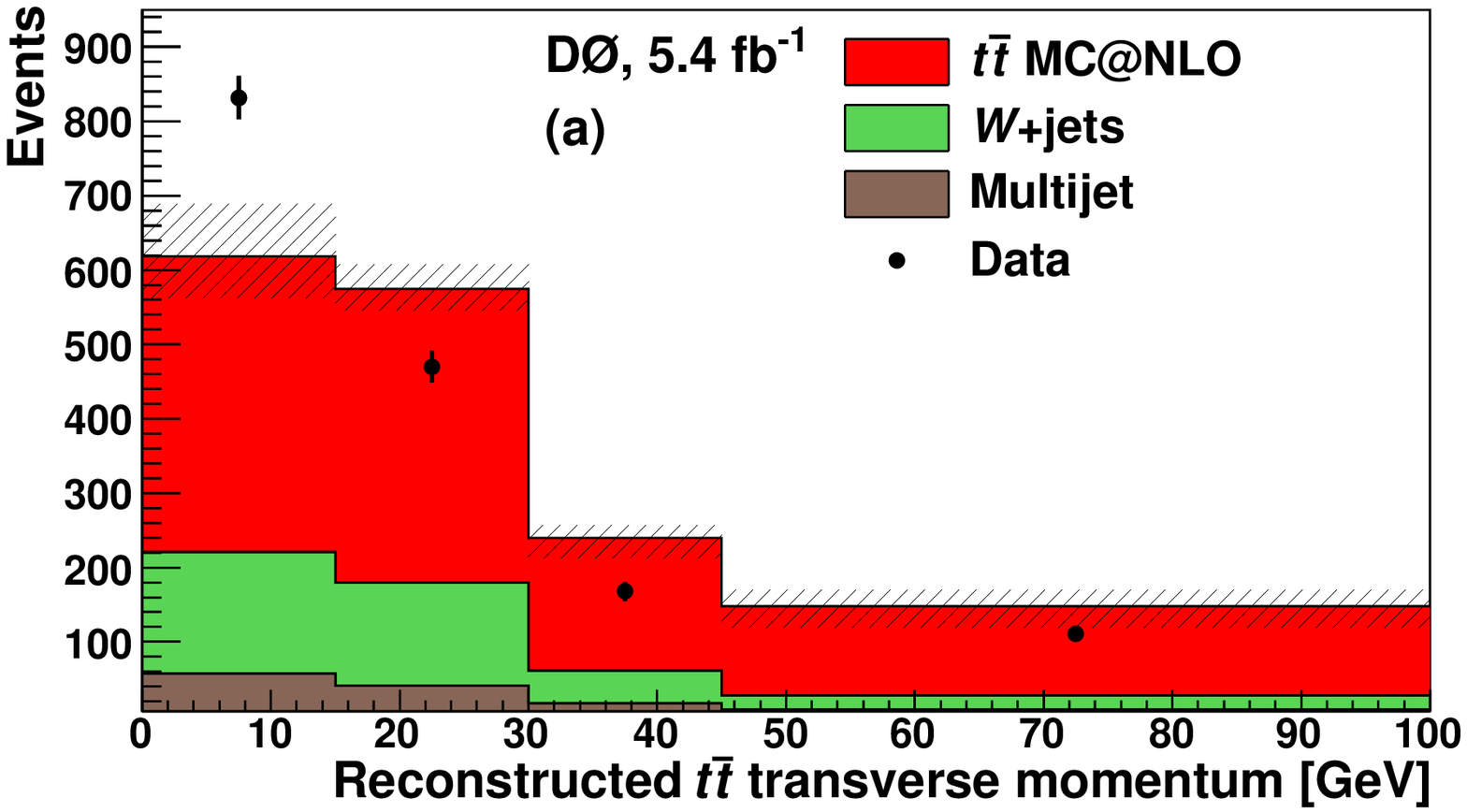}
\includegraphics[scale=.39]{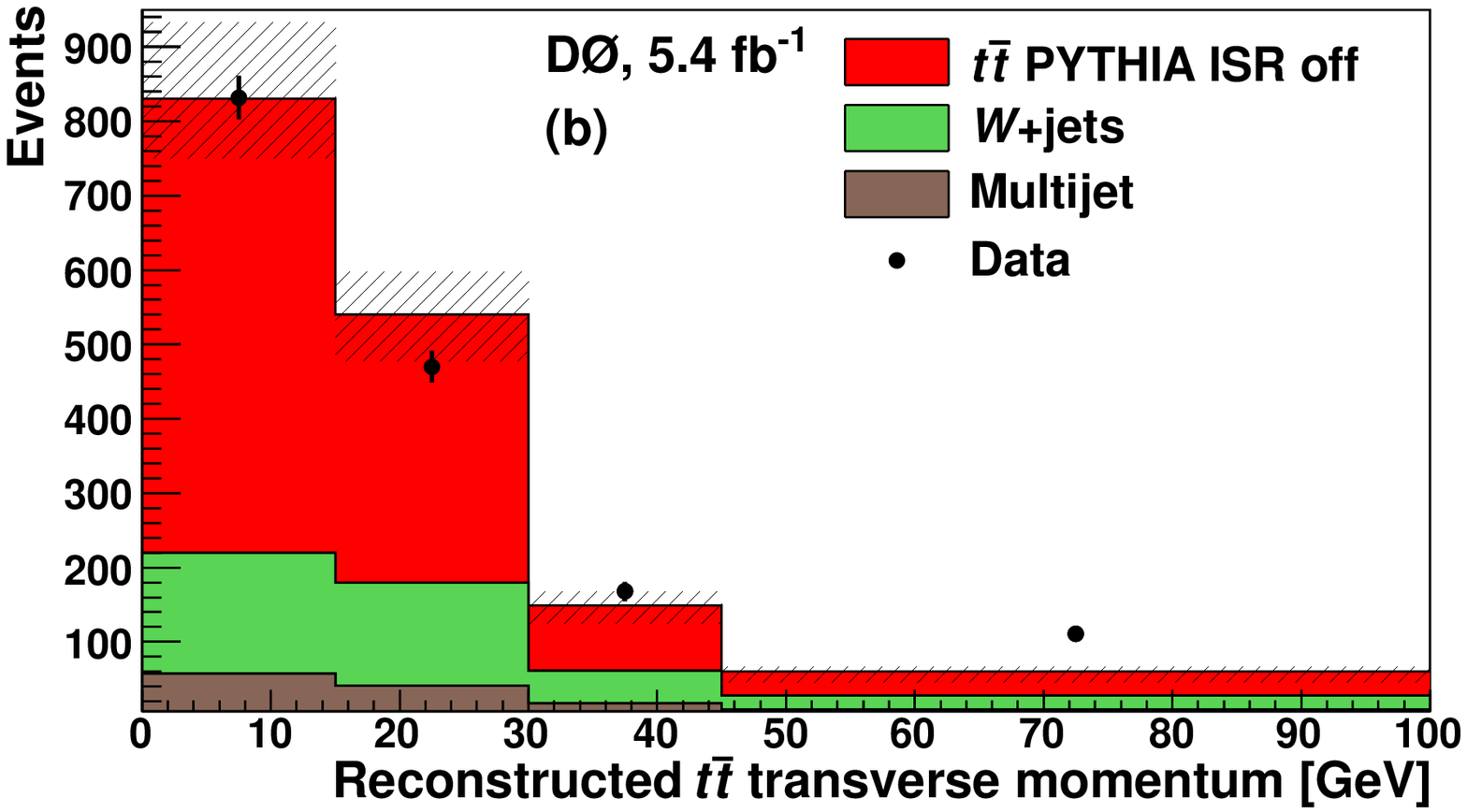}
\end{center}
\vspace{-0.6cm}
\caption{\label{fig:ttpt} The reconstructed \ttpt. 
The hatched band represents systematic uncertainties due to the jet energy scale and resolution. 
Data points are compared to predictions based on (a) \mcatnlo\ and (b) \pythia\ with ISR off. 
Bin widths correspond to about half of the detector resolution in \ttpt. }
\end{figure}

To further clarify this issue, dedicated measurements of \ttpt\ and detailed prediction for 
the dependence of \afb\ on this quantity are needed.
During the preparation of this paper,
the first such calculations became available~\cite{bib:revisit}.

%
%
\section{Summary}
We measure the forward-backward asymmetry in top quark-antiquark production,
defined according to the rapidity difference between the top and antitop quarks. 
After background subtraction, we find a reconstructed \ttbar\ asymmetry of 
$\afb = \left(9.2 \pm 3.7\right)$\%, to be compared
with the \mcatnlo-based prediction of $\left(2.4 \pm 0.7\right)$\%.
We find no statistically significant enhancements of \afb, neither for high \mttbar\ nor for large \absdy.

The reconstructed \ttbar\ asymmetry can be unfolded for acceptance and detector resolution. We apply two
unfolding procedures: a four-bin unfolding and an unfolding with fine binning and explicit regularization. 
We argue that the latter technique is better suited to estimate migration between the 
regions of positive and negative \dy\ and reduces the overall uncertainty on the unfolded result.
The asymmetry unfolded with fine binning and explicit regularization 
is $\left(19.6 \pm 6.5 \right)$\%, while \mcatnlo\ predicts a value of $\left(5.0 \pm 0.1 \right)$\%.

We also present an alternative approach that does not depend on a full reconstruction of the \ttbar\ system --- 
a measurement of a forward-backward asymmetry based only on the rapidity of
the lepton. To avoid large acceptance corrections, we use only the region $\absyl<1.5$.
We measure $\afbl = \left(14.2 \pm 3.8\right)$\% at the reconstruction level,
to be compared to the \mcatnlo-based prediction of $\left(0.8 \pm 0.6\right)$\%. 
Unfolding to the production level has a minimal effect on the lepton-based asymmetry,
and we find $\afbl = \left(15.2 \pm 4.0\right)$\% at the production level, 
to be compared with the \mcatnlo-based prediction of $\left(2.1 \pm 0.1\right)$\%.

The asymmetries measured in \DZ\ data disagree with the \mcatnlo-based predictions, with the 
most significant discrepancy above three SD. 
The \afb\ value measured at production level can also be compared to other 
SM calculations (e.g.~\cite{bib:kidonakis,bib:bernsi,bib:ahrens,bib:hollik}), 
which predict somewhat higher asymmetries.

%
%
\begin{acknowledgments}
We thank M.~Mangano, P.\,Z.~Skands, C.-P.~Yuan and L.~Dixon for enlightening discussions.
\input{acknowledgement.tex}
\end{acknowledgments}

\end{document}

%% file: author_list_for_ttbar_afb.tex
\affiliation{Universidad de Buenos Aires, Buenos Aires, Argentina}
\affiliation{LAFEX, Centro Brasileiro de Pesquisas F{\'\i}sicas, Rio de Janeiro, Brazil}
\affiliation{Universidade do Estado do Rio de Janeiro, Rio de Janeiro, Brazil}
\affiliation{Universidade Federal do ABC, Santo Andr\'e, Brazil}
\affiliation{Instituto de F\'{\i}sica Te\'orica, Universidade Estadual Paulista, S\~ao Paulo, Brazil}
\affiliation{Simon Fraser University, Vancouver, British Columbia, and York University, Toronto, Ontario, Canada}
\affiliation{University of Science and Technology of China, Hefei, People's Republic of China}
\affiliation{Universidad de los Andes, Bogot\'{a}, Colombia}
\affiliation{Charles University, Faculty of Mathematics and Physics, Center for Particle Physics, Prague, Czech Republic}
\affiliation{Czech Technical University in Prague, Prague, Czech Republic}
\affiliation{Center for Particle Physics, Institute of Physics, Academy of Sciences of the Czech Republic, Prague, Czech Republic}
\affiliation{Universidad San Francisco de Quito, Quito, Ecuador}
\affiliation{LPC, Universit\'e Blaise Pascal, CNRS/IN2P3, Clermont, France}
\affiliation{LPSC, Universit\'e Joseph Fourier Grenoble 1, CNRS/IN2P3, Institut National Polytechnique de Grenoble, Grenoble, France}
\affiliation{CPPM, Aix-Marseille Universit\'e, CNRS/IN2P3, Marseille, France}
\affiliation{LAL, Universit\'e Paris-Sud, CNRS/IN2P3, Orsay, France}
\affiliation{LPNHE, Universit\'es Paris VI and VII, CNRS/IN2P3, Paris, France}
\affiliation{CEA, Irfu, SPP, Saclay, France}
\affiliation{IPHC, Universit\'e de Strasbourg, CNRS/IN2P3, Strasbourg, France}
\affiliation{IPNL, Universit\'e Lyon 1, CNRS/IN2P3, Villeurbanne, France and Universit\'e de Lyon, Lyon, France}
\affiliation{III. Physikalisches Institut A, RWTH Aachen University, Aachen, Germany}
\affiliation{Physikalisches Institut, Universit{\"a}t Freiburg, Freiburg, Germany}
\affiliation{II. Physikalisches Institut, Georg-August-Universit{\"a}t G\"ottingen, G\"ottingen, Germany}
\affiliation{Institut f{\"u}r Physik, Universit{\"a}t Mainz, Mainz, Germany}
\affiliation{Ludwig-Maximilians-Universit{\"a}t M{\"u}nchen, M{\"u}nchen, Germany}
\affiliation{Fachbereich Physik, Bergische Universit{\"a}t Wuppertal, Wuppertal, Germany}
\affiliation{Panjab University, Chandigarh, India}
\affiliation{Delhi University, Delhi, India}
\affiliation{Tata Institute of Fundamental Research, Mumbai, India}
\affiliation{University College Dublin, Dublin, Ireland}
\affiliation{Korea Detector Laboratory, Korea University, Seoul, Korea}
\affiliation{CINVESTAV, Mexico City, Mexico}
\affiliation{Nikhef, Science Park, Amsterdam, the Netherlands}
\affiliation{Radboud University Nijmegen, Nijmegen, the Netherlands and Nikhef, Science Park, Amsterdam, the Netherlands}
\affiliation{Joint Institute for Nuclear Research, Dubna, Russia}
\affiliation{Institute for Theoretical and Experimental Physics, Moscow, Russia}
\affiliation{Moscow State University, Moscow, Russia}
\affiliation{Institute for High Energy Physics, Protvino, Russia}
\affiliation{Petersburg Nuclear Physics Institute, St. Petersburg, Russia}
\affiliation{Instituci\'{o} Catalana de Recerca i Estudis Avan\c{c}ats (ICREA) and Institut de F\'{i}sica d'Altes Energies (IFAE), Barcelona, Spain}
\affiliation{Stockholm University, Stockholm and Uppsala University, Uppsala, Sweden}
\affiliation{Lancaster University, Lancaster LA1 4YB, United Kingdom}
\affiliation{Imperial College London, London SW7 2AZ, United Kingdom}
\affiliation{The University of Manchester, Manchester M13 9PL, United Kingdom}
\affiliation{University of Arizona, Tucson, Arizona 85721, USA}
\affiliation{University of California Riverside, Riverside, California 92521, USA}
\affiliation{Florida State University, Tallahassee, Florida 32306, USA}
\affiliation{Fermi National Accelerator Laboratory, Batavia, Illinois 60510, USA}
\affiliation{University of Illinois at Chicago, Chicago, Illinois 60607, USA}
\affiliation{Northern Illinois University, DeKalb, Illinois 60115, USA}
\affiliation{Northwestern University, Evanston, Illinois 60208, USA}
\affiliation{Indiana University, Bloomington, Indiana 47405, USA}
\affiliation{Purdue University Calumet, Hammond, Indiana 46323, USA}
\affiliation{University of Notre Dame, Notre Dame, Indiana 46556, USA}
\affiliation{Iowa State University, Ames, Iowa 50011, USA}
\affiliation{University of Kansas, Lawrence, Kansas 66045, USA}
\affiliation{Kansas State University, Manhattan, Kansas 66506, USA}
\affiliation{Louisiana Tech University, Ruston, Louisiana 71272, USA}
\affiliation{Boston University, Boston, Massachusetts 02215, USA}
\affiliation{Northeastern University, Boston, Massachusetts 02115, USA}
\affiliation{University of Michigan, Ann Arbor, Michigan 48109, USA}
\affiliation{Michigan State University, East Lansing, Michigan 48824, USA}
\affiliation{University of Mississippi, University, Mississippi 38677, USA}
\affiliation{University of Nebraska, Lincoln, Nebraska 68588, USA}
\affiliation{Rutgers University, Piscataway, New Jersey 08855, USA}
\affiliation{Princeton University, Princeton, New Jersey 08544, USA}
\affiliation{State University of New York, Buffalo, New York 14260, USA}
\affiliation{Columbia University, New York, New York 10027, USA}
\affiliation{University of Rochester, Rochester, New York 14627, USA}
\affiliation{State University of New York, Stony Brook, New York 11794, USA}
\affiliation{Brookhaven National Laboratory, Upton, New York 11973, USA}
\affiliation{Langston University, Langston, Oklahoma 73050, USA}
\affiliation{University of Oklahoma, Norman, Oklahoma 73019, USA}
\affiliation{Oklahoma State University, Stillwater, Oklahoma 74078, USA}
\affiliation{Brown University, Providence, Rhode Island 02912, USA}
\affiliation{University of Texas, Arlington, Texas 76019, USA}
\affiliation{Southern Methodist University, Dallas, Texas 75275, USA}
\affiliation{Rice University, Houston, Texas 77005, USA}
\affiliation{University of Virginia, Charlottesville, Virginia 22901, USA}
\affiliation{University of Washington, Seattle, Washington 98195, USA}
\author{V.M.~Abazov} \affiliation{Joint Institute for Nuclear Research, Dubna, Russia}
\author{B.~Abbott} \affiliation{University of Oklahoma, Norman, Oklahoma 73019, USA}
\author{B.S.~Acharya} \affiliation{Tata Institute of Fundamental Research, Mumbai, India}
\author{M.~Adams} \affiliation{University of Illinois at Chicago, Chicago, Illinois 60607, USA}
\author{T.~Adams} \affiliation{Florida State University, Tallahassee, Florida 32306, USA}
\author{G.D.~Alexeev} \affiliation{Joint Institute for Nuclear Research, Dubna, Russia}
\author{G.~Alkhazov} \affiliation{Petersburg Nuclear Physics Institute, St. Petersburg, Russia}
\author{A.~Alton$^{a}$} \affiliation{University of Michigan, Ann Arbor, Michigan 48109, USA}
\author{G.~Alverson} \affiliation{Northeastern University, Boston, Massachusetts 02115, USA}
\author{G.A.~Alves} \affiliation{LAFEX, Centro Brasileiro de Pesquisas F{\'\i}sicas, Rio de Janeiro, Brazil}
% ...
%
%
\collaboration{The D0 Collaboration\footnote{with visitors from
$^{a}$Augustana College, Sioux Falls, SD, USA,
$^{b}$The University of Liverpool, Liverpool, UK,
$^{c}$SLAC, Menlo Park, CA, USA,
$^{d}$University College London, London, UK,
$^{e}$Centro de Investigacion en Computacion - IPN, Mexico City, Mexico,
$^{f}$ECFM, Universidad Autonoma de Sinaloa, Culiac\'an, Mexico,
and 
$^{g}$Universit{\"a}t Bern, Bern, Switzerland.
$^{\ddag}$Deceased.
}} \noaffiliation
\vskip 0.25cm

%% file: acknowledgement.tex
%
We thank the staffs at Fermilab and collaborating institutions,
and acknowledge support from the
DOE and NSF (USA);
CEA and CNRS/IN2P3 (France);
FASI, Rosatom and RFBR (Russia);
CNPq, FAPERJ, FAPESP and FUNDUNESP (Brazil);
DAE and DST (India);
Colciencias (Colombia);
CONACyT (Mexico);
KRF and KOSEF (Korea);
CONICET and UBACyT (Argentina);
FOM (The Netherlands);
STFC and the Royal Society (United Kingdom);
MSMT and GACR (Czech Republic);
CRC Program and NSERC (Canada);
BMBF and DFG (Germany);
SFI (Ireland);
The Swedish Research Council (Sweden);
and
CAS and CNSF (China).